\documentclass[10pt, twocolumn, twoside]{IEEEtran}

\ifCLASSINFOpdf
 \usepackage[pdftex]{graphicx}
\else
 \usepackage[dvips]{graphicx}
\fi
\usepackage{cite}
\usepackage{amsmath,amssymb,amsfonts,lipsum}
\usepackage{float}
\usepackage{graphicx}
\usepackage{textcomp}
\usepackage{color}
\usepackage{xcolor}
\usepackage{ragged2e}

\usepackage{algorithm,algpseudocode}
\usepackage{framed,multirow}
\usepackage{mwe}
\usepackage{diagbox}
\usepackage{mathtools}
\usepackage{stfloats}
\usepackage[caption=false, font=normalsize, labelfont=sf, textfont=sf]{subfig}
\def\BibTeX{{\rm B\kern-.05em{\sc i\kern-.025em b}\kern-.08em
		T\kern-.1667em\lower.7ex\hbox{E}\kern-.125emX}}

\begin{document}

\title{Exploring Evolutionary Spectral Clustering for Temporal-Smoothed \hspace{-3mm} Clustered \hspace{-3mm} Cell-Free \hspace{-3mm} Networking}

\author{Junyuan~Wang, \emph{Member,~IEEE}, Tianyao~Wu, Ouyang~Zhou, and Yaping~Zhu, \emph{Member,~IEEE}
\thanks{Manuscript received 29 August 2024; revised 6 November 2024; accepted 2 Decemberr 2024.
This work was supported in part by National Natural Science Foundation of China under Grants 62371344, 62001330 and 62101385, and in part by the Fundamental Research Funds for Central Universities.
The associate editor coordinating the review of this article and approving it for publication was Z. Becvar.
\textit{(Corresponding author: Yaping Zhu.)}}
\thanks{J. Wang, T. Wu, O. Zhou are with College of Electronic and Information Engineering, Tongji University, Shanghai, 201804, China (e-mail: \{junyuanwang, 2233010, zoy\}@tongji.edu.cn).}
\thanks{Y. Zhu is with School of Computer Science and Technology, Tongji University, Shanghai, 201804, China (e-mail: yapingzhu@tongji.edu.cn).}
}

\maketitle

\begin{abstract}
Clustered cell-free networking, which dynamically partitions the whole network into nonoverlapping subnetworks, has been recently proposed to mitigate the cell-edge problem in cellular networks. However, prior works only focused on optimizing clustered cell-free networking in static scenarios with fixed users. This could lead to a large number of handovers in the practical dynamic environment with moving users, seriously hindering the implementation of clustered cell-free networking in practice. This paper considers user mobility and aims to simultaneously maximize the sum rate and minimize the number of handovers. By transforming the multi-objective optimization problem into a time-varying graph partitioning problem and exploring evolutionary spectral clustering, a temporal-smoothed clustered cell-free networking algorithm is proposed, which is shown to be effective in  smoothing network partitions over time and reducing handovers while maintaining similar sum rate.
\end{abstract}

\begin{IEEEkeywords}
Clustered cell-free networking, user mobility, temporal smoothness,  handover, evolutionary spectral clustering
\end{IEEEkeywords}

\vspace{-6mm}
\section{Introduction}
\label{sec:introduction}
Current cellular network suffers from the well-known cell-edge problem, i.e., the users located close to the cell edge experience low data rates due to the strong interference from neighboring base-stations (BSs). Coordinating BSs to serve users was then studied, which originates from distributed antenna system \cite{Wang2015asymptotic} and is popularly known as cell-free massive multiple-input-multiple-output (MIMO) \cite{Ammar2021user,chen2023user,Zhang2024PilotAF}. 
However, coordinating a large number of BSs is impractical due to the high joint processing complexity and signaling overhead.
Clustered cell-free networking has been recently proposed in \cite{wang2023clustered,Dai2017Optimal} to dynamically partition the whole network into multiple independent operating subnetworks by clustering strongly interfered users and BSs into the same subnetwork. Thus, joint processing is only required within each subnetwork. This innovative clustered cell-free networking architecture amalgamates the advantages of the divide-and-conquer cellular network and cooperative transmission, making it a promising candidate for future ultra-dense wireless communication systems.

There have been a number of research works devoted to the optimization of clustered cell-free networking \cite{Dai2017Optimal,wang2023clustered,Yang2022C2,Xia2023complexity,Boxiang2024ISIT,chaowen2022CGN,Boxiang2024ICC}. 
Aiming at minimizing the joint processing complexity and signaling overhead, the number of subnetworks were maximized in \cite{Dai2017Optimal,wang2023clustered,Yang2022C2} given the total interference constraint \cite{Dai2017Optimal}, downlink per-user rate requirement \cite{wang2023clustered}, or uplink per-subnetwork capacity requirement \cite{Yang2022C2}. 
Other works aimed to maximize the uplink sum capacity of clustered cell-free networks \cite{Xia2023complexity,Boxiang2024ISIT}, the minimum subnetwork capacity \cite{chaowen2022CGN}, or the downlink sum rate \cite{Boxiang2024ICC}.
However, all the aforementioned works \cite{Dai2017Optimal,wang2023clustered,Yang2022C2,Xia2023complexity,Boxiang2024ISIT,chaowen2022CGN,Boxiang2024ICC} focus on static scenarios with fixed users, overlooking the dynamic nature of the real-world network where users are constantly moving and frequent update of clustered cell-free networking is required. The continuous changes in the association between users and BSs would result in a large number of handovers.\footnote{Note that handover has been extensively studied in cellular networks, e.g, \cite{DaCosta2018adaptive}. Yet handover is more challenging in clustered cell-free networks because both the BSs and users in a subnetwork could vary as users move.} It is therefore of paramount practical importance to consider the time-varying nature of the network and devise a resilient clustered cell-free networking scheme to user mobility.

In this paper, we focus on the clustered cell-free networking problem and consider the impact of user mobility. Specifically, we aim to maximize the sum rate and minimize the number of handovers. By defining temporal smoothness\footnote{Here, temporal smoothness describes the smoothness of the transition of clustered cell-free networking over time. Higher temporal smoothness indicates less changes in BS-user association and hence less handovers.} to reflect the number of handovers and modeling the network as a weighted undirected graph, we transform the multi-objective clustered cell-free networking problem into a  time-varying graph partitioning problem. We then explore evolutionary spectral clustering to solve it and propose a temporal-smoothed clustered cell-free networking algorithm. 
Simulation results demonstrate that in the context of user mobility,  the clustered cell-free networking result with the proposed algorithm can smoothly transit over time at the cost of little degradation in sum rate compared to the benchmark that maximizes the sum rate at each time instant independently. 
This validates the effectiveness of the proposed temporal-smoothed clustered cell-free networking algorithm in striking a favorable balance between reducing handovers and maximizing sum rate.

\vspace{-3mm}
\section{Network Model and Key Performance Indicators}
\label{sec:system model}
\subsection{Clustered Cell-Free Network Model}\label{subsec:system model}
Consider a large-scale clustered cell-free network which consists of $K$ users in $\mathcal{U}  =\left \{ u_{1} ,u_{2}, \cdots, u_{K} \right \}$ with $\left | \mathcal{U}  \right |=K$, and $L$ BSs in $\mathcal{B} =\left\{b_{1},b_{2},\cdots,b_{L}\right\}$ with $\left | \mathcal{B}  \right |=L $. All users and BSs are equipped with a single antenna each and randomly located throughout the area, and share the same time/frequency resources.
Different from the existing works that focused on static scenarios with fixed users \cite{Dai2017Optimal,wang2023clustered,Yang2022C2,Xia2023complexity,Boxiang2024ISIT,chaowen2022CGN,Boxiang2024ICC}, in this paper, we consider the impact of user mobility on clustered cell-free networking.
\footnote{In practical systems, a user might exit or be admitted to the network at any time. Here, we start from the assumption that a given set of users move in the network, while we will discuss later in Remark 1 that the random user joining and exiting issue can be implicitly handled by our proposed method.} 
Assume that the whole network is partitioned into $M$ nonoverlapping subnetworks, and the set of users and BSs in the $m$-th subnetwork at time $t$ is denoted as $\mathcal{C}_{m}^{(t)}$.
The network partition at time $t$ is then $\mathcal{M}^{(t)} =\left \{ \mathcal{C} _{1}^{(t)},\mathcal{C} _{2}^{(t)},\cdots,\mathcal{C} _{M}^{(t)} \right \}$, 
satisfying $\mathcal{C} _{m}^{(t)}\cap \mathcal{C} _{m^{'} }^{(t)}=\phi ,\forall m^{'}\ne m$ and $\bigcup_{m=1}^{M}\mathcal{C} _{m}^{(t)}=\mathcal{U}  \cup \mathcal{B}$
per the definition of clustered cell-free networking. 

\vspace{-3mm}
\subsection{Sum Rate}
Note that clustered cell-free networking is proposed with the aim of improving the achievable rates of the cell-edge users in current cellular networks by enabling joint processing of BSs in the same subnetwork. Therefore, a key performance indicator (KPI) considered in the existing works to measure the quality of network partition $\mathcal{M}^{(t)}$ is the sum rate of users. 

At time $t$, the sum rate $R^{(t)}(\mathcal{M})$ under any network partition $\mathcal{M}$ can be expressed as
\begin{equation}
\small
     R^{(t)} \left ( \mathcal{M}^{(t)}  \right ) = \sum_{m=1}^{M} \sum_{u_{k} \in \mathcal{C} _{m}^{(t)}} R_{k}^{(t)}\left(\mathcal{M}^{(t)}\right), \label{eq4}
\end{equation}
where $R_{k}^{(t)}\left(\mathcal{M}\right)$ is the achievable rate of user $u_{k}$  at time $t$ under partition $\mathcal{M}$, given by
\begin{equation}
\small
    R_{k}^{(t)}\left(\mathcal{M}\right) =\log_{2}{ \left ( 1+\mu _{k}^{(t)} \left(\mathcal{M}\right)\right )}, \label{eq5}
\end{equation}
where $\mu_{k}^{(t)}\left(\mathcal{M}\right)$ is the received signal-to-interference-plus-noise ratio (SINR) at user $u_{k}$. 
To decouple the networking problem and the specific spatial precoding scheme adopted in each subnetwork for the purpose of developing a general clustered cell-free networking algorithm that can be implemented along with any spatial precoding scheme, similar to \cite{wang2023clustered,Boxiang2024ICC}, we approximate the received SINR $\mu_{k}^{(t)}\left(\mathcal{M}\right)$ as 
\begin{equation}
\small
    \mu _{k}^{(t)}\left(\mathcal{M}\right)\approx \frac{P_{t}\left | h_{k,l_{k}^{*} }^{(t)} \right |^{2}  }{\sum_{b_{l} \notin \mathcal{C}_{m},b_{l} \in \mathcal{B}} {P_{t}\left | h_{k,l }^{(t)} \right |^{2} +\sigma ^{2}  } } ,\label{eq6}
\end{equation}
where $P_{t}$ is the transmit power of each BS. $\sigma ^{2}$ is the variance of the additive white Gaussian noise (AWGN). $h_{k,l}^{(t)}$ is the channel gain coefficient from BS $b_{l}$ to user $u_{k}$ at time $t$, and $b_{l_{k}^{*}}^{(t)}=\arg \max_{b_{l}\in \mathcal{B}}{|h_{k,l}^{(t)}|^2}$ is the best BS with the highest channel gain for user $u_{k}$ at time $t$.

\vspace{-3mm}
\subsection{Temporal Smoothness}
Clustered cell-free networking enables a flexible network architecture and thus the capability of maintaining low inter-subnetwork interference by dynamically clustering BSs and users into subnetworks. However, such flexibility could bring a large number of handovers, as maximizing the sum rate at each time instant independently could lead to significantly different network partitions at consecutive time instants due to user mobility. 
Such handovers incur additional transmission delay and signaling overhead, thus adversely degrading the system performance.  
To mitigate the frequent handover issue, as pointed out in \cite{chen2023user}, it is important to ensure a smooth transition of the network partition over time. 

Therefore, we propose to consider temporal smoothness as another KPI for clustered cell-free networking and define it as
\begin{equation}
\small
     S\left ( \mathcal{M}^{(t)}  \right ) \triangleq R^{(t-1)}\left ( \mathcal{M}^{(t)}  \right ). \label{eq7}
\end{equation}
The temporal smoothness $S\left(\mathcal{M}^{(t)}\right)$ is defined based on the fact that for two possible partitions that are equally effective in maximizing the sum rate at time $t$, the one similar to the network partition at time $t-1$ is preferred as it exhibits better adaptability to the preceding partition 
and requires fewer handovers. Since the network partition at time $t-1$ maximizes the sum rate at that time, the similarity between the network partition $\mathcal{M}^{(t)}$ at time $t$ and $\mathcal{M}^{(t-1)}$ at time $t-1$, i.e., temporal smoothness, can be measured by $R^{(t-1)}\left ( \mathcal{M}^{(t)}\right )$. The higher the $R^{(t-1)}\left ( \mathcal{M}^{(t)}\right )$, the smoother the transition of the clustered cell-free network partition from time $t-1$ to time $t$.\footnote{The effectiveness of the defined temporal smoothness in reflecting the number of handovers will be demonstrated later in Fig. 1 in Section IV.}

\section{Temporal-Smoothed Clustered Cell-Free Networking}

In contrast to previous studies \cite{Dai2017Optimal,wang2023clustered,Yang2022C2,Xia2023complexity,Boxiang2024ISIT,chaowen2022CGN,Boxiang2024ICC}  that optimized clustered cell-free networking at each time instant independently, this paper aims to maximize both the sum rate and the temporal smoothness of the network partition, and proposes a temporal-smoothed clustered cell-free networking algorithm.

The multi-objective temporal-smoothed clustered cell-free networking problem is formulated as

\vspace{-3mm}
{\small
\begin{align}
\left ( \mathcal{P}_1  \right ) ~
\max_{\mathcal{M}^{(t)}} 
\quad & {R^{(t)}\left ( \mathcal{M}^{(t)}\right )} \label{eq8}  \\ 
\max_{\mathcal{M}^{(t)}} 
\quad & {S\left( \mathcal{M}^{(t)}\right )}\label{eq9}& \\
\mbox{s.t.} \quad 
&\bigcup_{m=1}^{M} \mathcal{C}_{m}^{(t)}=\mathcal{U} \cup \mathcal{B} , \label{eq10} & \\
& \mathcal{C}_{m}^{(t)}\cap \mathcal{C}_{m^{'} }^{(t)}=\phi ,\forall m^{'}\ne m \label{eq11}, & \\
&  \mathcal{C}_{m}^{(t)} \cap \mathcal{B} \ne \phi, \forall m ,\label{eq12} &
\end{align}} 
where \eqref{eq10} and \eqref{eq11} follow the principle of clustered cell-free networking. Constraint \eqref{eq12} ensures that at least one BS exists in each subnetwork to provide services to users therein.\footnote{Note that a subnetwork can contain BSs only. In this case, these BSs are switched into sleep mode for energy saving.}

\vspace{-3mm}
\subsection{Problem Reformulation}\label{subsec:Problem Reformulation}
Note that problem $\mathcal{P}_1$ is challenging to solve due to the non-convex objective functions of combinatorial optimization variables. Inspired by \cite{wang2023clustered,Dai2017Optimal}, let us first transform $\mathcal{P}_1$ into a multi-objective graph partitioning problem. Specifically, we model the network as a weighted undirected time-varying graph $\mathcal{G}^{(t)}  =\left ( \mathcal{V}^{(t)} ,\mathcal{E}^{(t)} ,\mathbf{W}^{(t)}\right)$, where $\mathcal{V}^{(t)} =\left\{v_1^{(t)},v_2^{(t)},\cdots,v_L^{(t)}\right \} $ is the vertex set at time $t$. Each vertex $v_i^{(t)}$ contains BS $b_i$ and the users whose best BS with the highest channel gain is BS $b_{i}$, i.e., 
\begin{equation}
\small
    v_{i}^{(t)}=\left\{b_{i} \cup \left \{ u_{k} : b_{l_{k}^{*}}=b_i,\forall u_{k} \in \mathcal{U}  \right \} \right\}\label{eq13}.
\end{equation}
$\mathcal{E}^{(t)} =\left \{ \left ( v_i^{(t)},v_j^{(t)} \right ) :\forall v_i^{(t)},v_j^{(t)}\in\mathcal{V}^{(t)}  \right \} $ denotes the edge set, and
each edge $\left ( v_i^{(t)},v_j^{(t)} \right )$ has a weight $w_{i,j}^{(t)}$ given by 
\begin{equation}
\small
    w_{i,j}^{(t)}=
    \begin{cases}
    \sum_{u_{k} \in v_{i}^{(t)} } \hspace{-1mm}\frac{\left | h_{k,j}^{(t)} \right | ^2}{\left | h_{k,i}^{(t)} \right | ^2} \hspace{-0.5mm} + \hspace{-0.5mm}\sum_{u_{k} \in v_{j}^{(t)}} \hspace{-1mm}\frac{\left | h_{k,i}^{(t)} \right | ^2}{\left | h_{k,j}^{(t)} \right | ^2},\hspace{-2.5mm}& \text{if $i \ne j$;}\\
    0,\hspace{-2.5mm}& \text{otherwise.} \label{eq14}
    \end{cases}  
\end{equation}
$\mathbf{W}^{(t)}=[w_{i,j}^{(t)}] \in \mathbb{R} ^{L\times L}$ is the weight matrix.
Each subnetwork can be then represented by a subgraph of $\mathcal{G}^{(t)}$. Specifically, for the $m$-th subgraph corresponding to the $m$-th subnetwork at time $t$, its vertex set is denoted as $\mathcal{\hat{C}}_{m}^{(t)}=\{v_{i}^{(t)}:b_{i}\in \mathcal{C}_{m}^{(t)}\}$.

\noindent \textbf{Remark 1:} In practical systems, an existing user may leave the network at the next time instant and new users might be admitted to the network. The above graphical model of the network is a unified model that can well accommodate such scenarios involving user joining and exiting by simply updating the users included in each vertex  at each time $t$ along with the weight matrix $\mathbf{W}^{(t)}$. 
Therefore, although we initially assumed that the users admitted to the network remain the same, the situations with random user joining and/or exiting can be well handled by modeling the network as graph $\mathcal{G}^{(t)}$ and formalize the original clustered cell-free networking problem $\mathcal{P}_{1}$ as a graph partitioning problem.

Following the approach in \cite{wang2023clustered}, it can be found that maximizing the sum rate can be transformed to minimizing the sum cut $\sum_{m=1}^{M} \text{cut}^{(t)}(\mathcal{\hat{C}}_{m}^{(t)})$ of  partition $\mathcal{\hat{M}}^{(t)}=\{\hat{\mathcal{C}}^{(t)}_{1},\hat{\mathcal{C}}^{(t)}_{2}, \cdots, \hat{\mathcal{C}}^{(t)}_{M}$\} on graph $\mathcal{G}^{(t)}$. The cut function $\text{cut}^{(t^{'})}(\mathcal{\hat{C}}_{m}^{(t)})$ is defined as the sum weight of the edges that connect the vertices inside and outside $\mathcal{\hat{C}}_{m}^{(t)}$ on graph $\mathcal{G}^{(t^{'})}$, given by
\begin{equation}
\small
    \text{cut}^{(t^{'})}(\mathcal{\hat{C}}_{m}^{(t)})=\sum _{v_{i}^{(t)}\in \mathcal{\hat{C}}_{m}^{(t)}}\sum _{v_{j}^{(t)}\notin {\mathcal{\hat{C}}}_{m}^{(t)}} w_{i,j}^{(t^{'})}. \label{eq16}
\end{equation}
The temporal-smoothed clustered cell-free networking problem $\mathcal{P}_1$ given in \eqref{eq8}--\eqref{eq12} can be then transformed to  

\vspace{-4mm}
{\small
\begin{align}
\left ( \mathcal{P}_2  \right )~ 
\min_{\mathcal{\hat{M}}^{(t)}} 
\quad & \sum_{m=1}^{M}{\text{cut}^{(t)}}(\mathcal{\hat{C}}_{m}^{(t)}) \label{eq17}  \\ 
\min_{\mathcal{\hat{M}}^{(t)}} 
\quad & \sum_{m=1}^{M}{\text{cut}^{(t-1)}}(\mathcal{\hat{C}}_{m}^{(t)}) \label{eq18}& \\
\mbox{s.t.} \quad 
& \bigcup_{m=1}^{M} \mathcal{\hat{C}}_{m}^{(t)} = \mathcal{V}^{(t)} , \label{eq19} & \\
& \hspace{-0.5mm}\mathcal{\hat{C}}_{m}^{(t)}\cap  \mathcal{\hat{C}}_{m^{'} }^{(t)} =\phi, \forall m^{'}\ne m \label{eq20}. & 
\end{align}}

Clearly, $\mathcal{P}_2$ is a multi-objective optimization problem of graph partitioning, where the first objective in \eqref{eq17} aims to minimize the sum cut on graph $\mathcal{G}^{(t)}$ at the current time $t$, and the second objective in \eqref{eq18} 
aims to minimize the sum cut of the current partition $\mathcal{\hat{M}}^{(t)}$ on graph $\mathcal{G}^{(t-1)}$. 
Therefore, \eqref{eq17} measures how well partition $\mathcal{\hat{M}}^{(t)}$ fits the current graph $\mathcal{G}^{(t)}$ at time $t$, and \eqref{eq18} measures how smooth the transition from partition $\mathcal{\hat{M}}^{(t-1)}$ at time $t-1$ to $\mathcal{\hat{M}}^{(t)}$ at time $t$ is. Finding an optimal graph partition that simultaneously minimizes the sum cut and maximizes the temporal smoothness is usually impossible. This is because the optimal partition that maximizes the sum rate is different from one time instant to the next due to user movement, while maximizing the temporal smoothness leads to unchanged partition over time. 

To tackle the challenge of the multiple objectives in $\mathcal{P}_{2}$, we transform it into a single-objective optimization problem by using a linear weighting method. As such, existing single-objective optimization techniques can be leveraged to obtain a set of solutions that provide different trade-offs between the two objectives. Specifically, problem $ \mathcal{P}_2 $ is rewritten as the following time-varying graph partitioning problem

\vspace{-3mm}
{\small
\begin{align}
    \left ( \mathcal{P}_3  \right )~ 
    \min_{\mathcal{\hat{M}}^{(t)}} 
    ~ & \alpha\hspace{-1mm}\sum_{m=1}^{M}\hspace{-1mm}{\text{cut}^{(t)}}(\mathcal{\hat{C}}_{m}^{(t)}) {+} (1{-}\alpha)\hspace{-1mm}\sum_{m=1}^{M}\hspace{-1mm}{\text{cut}^{(t-1)}}(\mathcal{\hat{C}}_{m}^{(t)}) \label{eq21}  \\ 
    \mbox{s.t.} \quad &\eqref{eq19}, \eqref{eq20},
    & 
\label{eq23} 
\end{align}}where $\alpha \in \left [ 0,1 \right ] $ is the weighting coefficient, reflecting the trade-off between sum cut and temporal smoothness.

\vspace{-1mm}
\subsection{Temporal-Smoothed Clustered Cell-Free Networking}\label{subsec:Algorithm}
This section explores the evolutionary spectral clustering framework in \cite{ chi2007evolutionary} to solve problem $\mathcal{P}_3$. Specifically, let us define an indicator matrix $\mathbf{Z}^{(t)} \in \mathbb{R} ^{L\times M} $ whose $i$th row and $m$th column element $z_{i,m}^{(t)}$ is given by 
\begin{equation}
\small
    z_{i,m}^{(t)}=
    \begin{cases}
        1,& \text{if $ {v}_{i}^{(t)} \in \mathcal{\hat{C}}_m^{(t)}$;}\\
        0,& \text{if $ {v}_{i}^{(t)} \notin \mathcal{\hat{C}}_m^{(t)}$,}
    \end{cases} \label{eq24}
\end{equation}
for any ${v}_{i}^{(t)} \in \mathcal{V}^{(t)},i=1,2,\cdots,L$, and $m=1,2,\cdots,M$. 
Given the indicator matrix $\mathbf{Z}^{(t)}$, the sum cut functions in \eqref{eq17} and \eqref{eq18} can be equivalently written as \cite{von2007tutorial}
\begin{equation}
\small
    \sum_{m=1}^{M}{\text{cut}^{(t)}}(\mathcal{\hat{C}}_{m}^{(t)})=\mathrm{Tr} \left ( {\mathbf{Z}^{(t)} }^{T} \mathbf{L}^{(t)} \mathbf{Z}^{(t)}  \right ),  
    \label{eq25}
\end{equation}
\begin{equation}
\small
    \sum_{m=1}^{M}{\text{cut}^{(t-1)}}(\mathcal{\hat{C}}_{m}^{(t)})=\mathrm{Tr} \left ( {\mathbf{Z}^{(t)} }^{T} \mathbf{L}^{(t-1)} \mathbf{Z}^{(t)}  \right ), 
    \label{eq26}
\end{equation}
where $\mathbf{L}^{(t)}$ is the Laplacian matrix of graph $\mathcal{G}^{(t)}$, given by
\begin{equation}
\small
    \mathbf{L}^{(t)}=\mathbf{D}^{(t)}-\mathbf{W}^{(t)}.\label{eq29}
\end{equation}
$\mathbf{D}^{(t)}=\text{diag}\left(d_1^{(t)},d_2^{(t)},\cdots,d_L^{(t)}\right)$ denotes the degree matrix at time $t$, where $d_i^{(t)}=\sum_{j=1}^{L} {w}_{i,j}^{(t)}$ is the degree of vertex ${v}_i^{(t)}$. 

By substituting \eqref{eq25} and \eqref{eq26} into \eqref{eq21}, we can rewrite the objective function as
\begin{equation}
\small
    \begin{split}
     & \min_{\mathbf{Z}^{(t)}} \mathrm{Tr} \left [ {\mathbf{Z}^{(t)} }^{T} \left(\alpha \mathbf{L}^{(t)}+ (1-\alpha )  \mathbf{L}^{(t-1)}\right) \mathbf{Z}^{(t)}  \right ]
     \label{eq30}.        
    \end{split}
\end{equation}
According to \eqref{eq24}, constraint \eqref{eq19} can be transformed to
\begin{equation}
    \small \mathrm{Tr} \left ( {\mathbf{Z}^{(t)} }^{T}  \mathbf{Z}^{(t)}  \right ) =L. \label{eq31}
\end{equation}
We can then relax the graph partitioning problem $\mathcal{P}_3$ in \eqref{eq21}--\eqref{eq23} into an evolutionary spectral clustering problem as

\vspace{-3mm}
{\small
\begin{align}
    \left ( \mathcal{P}_4  \right ) \;
    \min_{\mathbf{Z}^{(t)}}&~ \mathrm{Tr} \left [ {\mathbf{Z}^{(t)} }^{T} \left(\alpha \mathbf{L}^{(t)}+ (1-\alpha )  \mathbf{L}^{(t-1)}\right) \mathbf{Z}^{(t)}  \right ]  
    \label{eq32}\\
    \text {s.t.}&~ \mathrm{Tr} \left ( {\mathbf{Z}^{(t)} }^{T}  \mathbf{Z}^{(t)}  \right ) =L, \label{eq33} 
\end{align}}for which, constraint \eqref{eq23} is implicitly satisfied as the clustering algorithms in data science aim to cluster data points into disjoint groups.
The optimal solution $\mathbf{Z}^{(t)}_{\star}$ to this trace minimization problem $\mathcal{P}_4$ can be obtained according to \cite{von2007tutorial} by first selecting the eigenvectors corresponding to the $M$ smallest eigenvalues of matrix $\alpha \mathbf{L}^{(t)}+ (1-\alpha )  \mathbf{L}^{(t-1)}$ as the column vectors of $\mathbf{Y}^{(t)}$, after which, applying \textit{k}-means clustering algorithm to the row vectors  $\mathbf{y}_{1}^{r},\mathbf{y}_{2}^{r},\cdots,\mathbf{y}_{L}^{r}$ of $\mathbf{Y}^{(t)}$ yields graph partition $\mathcal{\hat{M}}^{(t)}_{\star}=\{\hat{\mathcal{C} }^{(t)}_{\star1},\hat{\mathcal{C} }^{(t)}_{\star2}, \cdots, \hat{\mathcal{C} }^{(t)}_{\star M}\}$ at time $t$. The optimal network partition $\mathcal{M}_{\star}^{(t)} =
\left \{ \mathcal{C}_{\star1}^{(t)}, \mathcal{C}_{\star 2}^{(t)}, \cdots, \mathcal{C}_{\star M}^{(t)}  \right \}$ is then obtained from \eqref{eq13} with
\begin{align}
\small
    \mathcal{C}_{\star m}^{(t)}=&\left\{b_{i}: {v} _{i}^{(t)} \in \hat{C}_{\star m}^{(t)}, i= 1,2,\cdots,L\right\} \nonumber \\
    &\cup \left \{u_{k}: u_{k} \in v_{i}^{(t)}, \forall u_{k} \in \mathcal{U}, \forall {v} _{i}^{(t)} \in \hat{C}_{\star m}^{(t)}\right\}.\label{eq34}
\end{align}
The details of the proposed temporal-smoothed clustered cell-free networking algorithm are summarized as Algorithm 1.

\vspace{-2mm}
\subsection{Complexity Analysis}
The proposed temporal-smoothed clustered cell-free networking algorithm is based on evolutionary spectral clustering, for which the computational complexity primarily depends on the eigenvalue decomposition of the modified Laplacian matrix with cubic complexity. As the modified Laplacian matrix $\alpha \mathbf{L}^{(t)}+ (1-\alpha)\mathbf{L}^{(t-1)}$ in our algorithm is an $L$-dimensional matrix, the computational complexity of the proposed temporal-smoothed clustered cell-free networking algorithm is $O\left(L^3\right)$.

\begin{algorithm}[t]
    \caption{Temporal-smoothed clustered cell-free networking}
    \label{alg1}
    \begin{algorithmic}[1]
        \Require Weight matrices $\mathbf{W}^{(t-1)}$ and $\mathbf{W}^{(t)}$, coefficient $\alpha$.
        \State Compute $\mathbf{L}^{(t-1)}$ and $\mathbf{L}^{(t)}$ according \eqref{eq29};
        \State Compute $M$ eigenvectors corresponding to the $M$ smallest eigenvalues of $\alpha \mathbf{L}^{(t)}+ (1-\alpha )  \mathbf{L}^{(t-1)}$; 
        \State Construct $\mathbf{Y}^{(t)}$ by collecting the $M$ eigenvectors as column vectors;
        \State Run the \textit{k}-means algorithm to cluster row vectors of $\mathbf{Y}^{(t)}$ into $M$ clusters and obtain $\hat{\mathcal{C}}^{(t)}_{\star 1},\hat{\mathcal{C}}^{(t)}_{\star2},\cdots,\hat{\mathcal{C}}^{(t)}_{\star M} $;
        \State Obtain $\mathcal{M}_{\star}^{(t)} =\left \{ \mathcal{C}_{\star1}^{(t)},\mathcal{C}_{\star2}^{(t)},\cdots,\mathcal{C}_{\star M}^{(t)}  \right \}$ according to \eqref{eq34};
        \Ensure Network partition $\mathcal{M}_{\star}^{(t)}$.
    \end{algorithmic}
\end{algorithm}

\section{Simulation Results}
In this section, we evaluate the performance of the proposed temporal-smoothed clustered cell-free networking algorithm and present the simulation results. Following the system model in Section \ref{subsec:system model}, $K$ users and $L$ BSs are randomly distributed within a $1\times1$ area. Assume that each user moves following the random waypoint model \cite{Lin2013Towards}. At any time $t$, each user at any waypoint randomly selects a transition length within $\left [ 0, 0.5 \right ]$ and a direction $\theta$ within $\left [ 0, 2\pi  \right ]$, and then moves to the next waypoint at time $t+1$. 
The channel gain coefficient is modeled by 
$h_{k,l}=d_{k,l}^{-{\beta}/{2}} \cdot g_{k,l}$, where $d_{k,l}$ is the Euclidean distance between user $u_{k}$ and BS $b_{l}$, and $\beta$ is the path-loss exponent. $g_{k,l} \sim \mathcal{CN}{(0,1)}$ denotes the small-scale fading coefficient with zero mean and unit variance.
To avoid the impact of fast-varying small-scale fading on the clustered cell-free networking result, we ignore the small-scale fading when performing the clustered cell-free networking algorithm. 
For downlink data transmission, we apply zero-forcing beamforming (ZFBF) in each subnetwork to eliminate the intra-subnetwork interference. As ZFBF requires the number of BSs in a subnetwork to be no less than the number of users, a user in the subnetwork with more users than BSs has zero rate.  

\begin{figure}[t]
    \centering
    \includegraphics[width=0.38\textwidth]{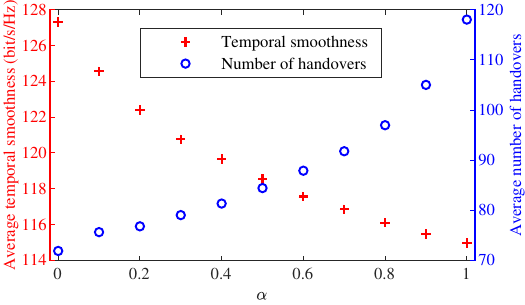}
    \caption{Average temporal smoothness and average number of handovers versus weighting coefficient $\alpha$. $K=30, L=50, M=20,\beta = 4, P_{t}/\sigma^{2}=0 \mathrm{dB}$.}
    \label{fig4}
\vspace{-5mm}
\end{figure}

To demonstrate the performance of the proposed algorithm, we adopt the recently proposed algorithm in \cite{wang2023clustered}, which decomposes the network based on the criterion of maximizing the sum rate only and runs independently at each time instant.  
This algorithm is a special case of our proposed clustered cell-free networking scheme with $\alpha=1$ and provides an upper-bound of the sum rate performance. 
The number of handovers is characterized by the number of new connections established at time $t$ compared to those at time $t-1$.
To mitigate the effect of the random layout of users and BSs, we conduct 5000 random realizations of the network layout and obtain the average sum rate and average number of handovers.

To show the rationality of the defined temporal smoothness in \eqref{eq7} in reflecting the number of required handovers, Fig.~\ref{fig4} presents both the average temporal smoothness and the average number of handovers with our proposed temporal-smoothed clustered cell-free networking algorithm by varying weight coefficient $\alpha$.
As can be seen from Fig.~\ref{fig4}, by increasing $\alpha$, the average temporal smoothness decreases monotonically and the average number of handovers increases monotonically. This indicates that higher temporal smoothness consequently leads to a lower number of handovers, validating the effectiveness of maximizing temporal smoothness of clustered cell-free networking for minimizing the number of handovers.

\begin{figure}[t]
    \centering
    \subfloat[at time $t-1$]{
        \label{fig1a}
        \includegraphics[width=0.24\textwidth,height=1.1in]{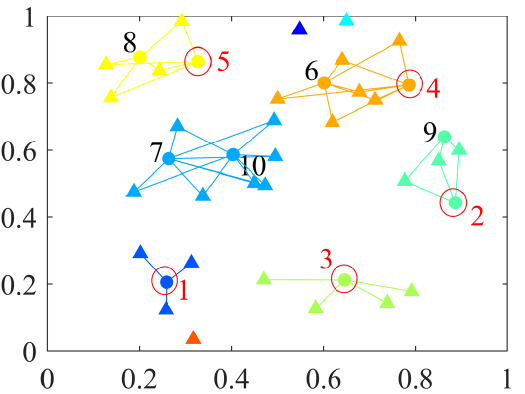}} \\
    \subfloat[at time $t$, $\alpha=0.5$]{
        \label{fig1b}
        \includegraphics[width=0.24\textwidth,height=1.1in]{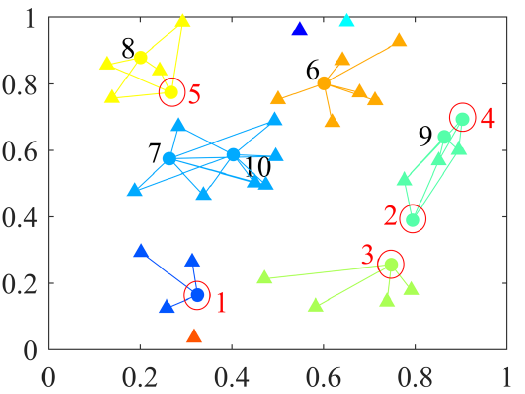}}
    \subfloat[at time $t$, Algorithm in \cite{wang2023clustered}]{
        \label{fig1c}
        \includegraphics[width=0.24\textwidth,height=1.1in]{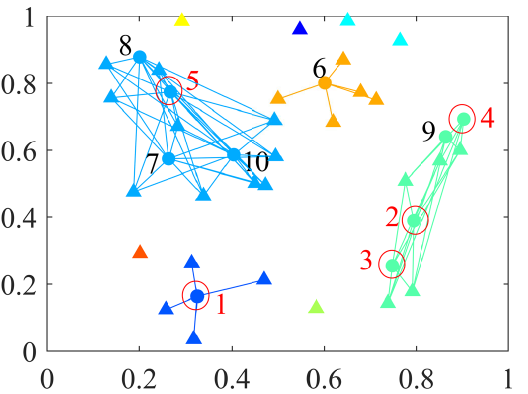}}
    \caption{Snapshots of the network partition of a randomly generated network (a) at time $t-1$, (b) at time $t$ with the proposed temporal-smoothed clustered cell-free networking algorithm when $\alpha=0.5$ and (c) at time $t$ with the algorithm in \cite{wang2023clustered}. Triangles and circles represent BSs and users, respectively, and different subnetworks are shown in different colors. $K=10, L=30, M=9, \beta = 4$.}
    \label{fig1}
\vspace{-3.5mm}
\end{figure}

Fig.~\ref{fig1} illustrates the networking results of a randomly generated topology by adopting the proposed temporal-smoothed clustered cell-free networking algorithm with $\alpha=0.5$ and the benchmarking algorithm in \cite{wang2023clustered}. For the sake of comparison, the network partition at preceding time $t-1$ is also shown. From time $t-1$ to  $t$, users 1-5 move to new locations.
It can be observed from Fig.~\ref{fig1} that both algorithms successfully partition the network into non-overlapping subnetworks, where users and BSs in the same color represent that they are in the same subnetwork. 
By comparing Fig.~\ref{fig1}\subref{fig1b} and Fig.~\ref{fig1}\subref{fig1c} against Fig.~\ref{fig1}\subref{fig1a}, we can see that the network partitioning result in Fig.~\ref{fig1}\subref{fig1c} is significantly different from that in Fig.~\ref{fig1}\subref{fig1a} due to the movement of users 1-5, while the users in Fig.~\ref{fig1}\subref{fig1b} still try to maintain the networking result at time $t-1$. For instance, although user 1 and user 3 move, they remain in the preceding subnetwork with our proposed algorithm, requiring no handovers. The results demonstrate that our algorithm can yield temporal-smoothed network partitions over time, leading to a considerable reduction in the number of handovers.

\begin{figure}[t]
    \centering
    \subfloat[$L=50$]{
        \label{fig5a}
        \includegraphics[width=0.25\textwidth,height=1.25in]{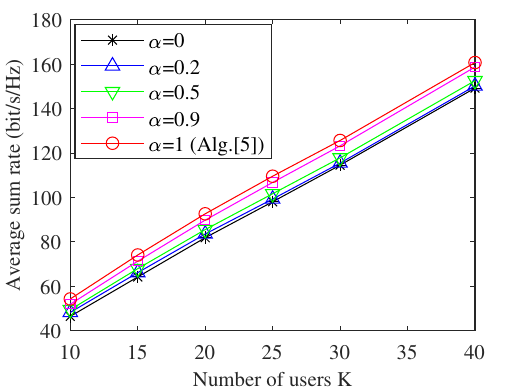}}
    \subfloat[$L=50$]{
        \label{fig5b}
        \includegraphics[width=0.25\textwidth,height=1.25in]{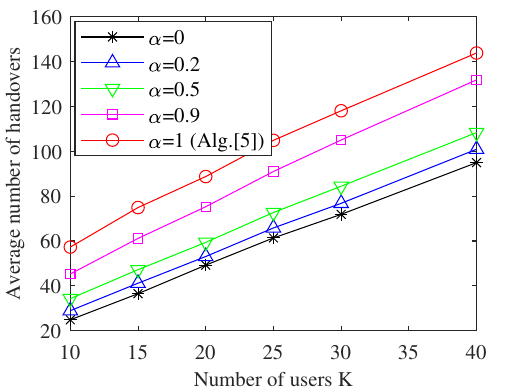}}
    
    \subfloat[$K=100$]{
        \label{fig5c}
        \includegraphics[width=0.25\textwidth,height=1.25in]{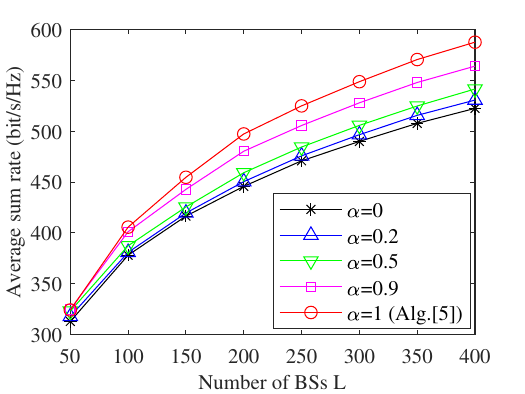}}
    \subfloat[$K=100$]{
        \label{fig5d}
        \includegraphics[width=0.25\textwidth,height=1.25in]{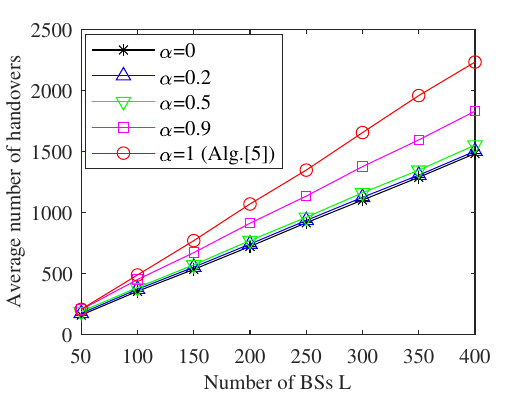}}
    \caption{(a) (c) Average sum rate and (b) (d) average number of handovers with the proposed temporal-smoothed clustered cell-free networking algorithm and the algorithm in \cite{wang2023clustered}. $M=20,\beta = 4, P_{t}/\sigma^{2}=0 \mathrm{dB}$.}
    \label{fig5}
    \vspace{-4mm}
\end{figure}

Fig.~\ref{fig5} further presents the average sum rate and the average number of handovers with the proposed temporal-smoothed algorithm and the algorithm in \cite{wang2023clustered}.
We can see from Fig.~\ref{fig5} that as the weighting coefficient $\alpha$ decreases, both the average sum rate and the average number of handovers with our proposed algorithm decrease. This is due to the fact that with a lower $\alpha$, the proposed algorithm prioritizes to maintain the preceding networking result at the cost of sum rate degradation. That is, our algorithm provides a trade-off between the sum rate performance and handover cost by fine-tuning the weighting coefficient $\alpha$. However, the reduction in the number of handovers becomes more pronounced than the reduction in sum rate as $\alpha$ increases. This indicates that by choosing a larger $\alpha$, the number of handovers can be significantly reduced with the proposed temporal-smoothed clustered cell-free networking algorithm yet at little sum rate loss.
For instance, when the number of users $K=30$, the number of BSs $L=50$ and $\alpha=0.9$, the average number of handovers with the proposed algorithm is 11.0\% less than that with the benchmark in \cite{wang2023clustered}, while only 1.9\% degradation in sum rate is observed.

\vspace{-1mm}
\section{Conclusion}
This paper considered user mobility in wireless networks and studied the clustered cell-free networking problem with the objective of jointly maximizing the sum rate and minimizing the number of handovers. By defining the temporal smoothness of network partition to reflect the required number of handovers and transforming the multi-objective optimization problem into a time-varying graph partitioning problem, a novel temporal-smoothed clustered cell-free networking algorithm is proposed based on evolutionary spectral clustering. Simulation results validate its effectiveness in smoothing the time-varying networking result, and show that the proposed algorithm can efficiently reduce the handovers while maintaining similar sum rate performance to the benchmark. In the future, the temporal-smoothed clustered cell-free working problem will be further studied by maximizing the minimum rate of users to improve user fairness or considering the specific service quality requirements of users. 

\bibliographystyle{IEEEtran}
\bibliography{IEEEabrv,Mybib}
 
\end{document}